%% file: kara_manunscript_accepted.tex
\documentclass{elektr}
\usepackage{hyperref}
\hypersetup{
colorlinks=true,
urlcolor=blue,
citecolor=blue}
\usepackage[all]{xy,xypic}
\usepackage{amsfonts,amssymb,amsmath,amsgen,amsopn,amsbsy,theorem,graphicx,epsfig}
\usepackage{eufrak,amscd,bezier,latexsym,mathrsfs,eurosym,enumerate}
\usepackage{xfrac}
\usepackage{epstopdf}
\usepackage[utf8]{inputenc}
\usepackage[english]{babel}
\usepackage{cleveref,multirow}
\usepackage[dvipsnames]{xcolor}
\usepackage[pagewise]{lineno}
\usepackage{graphics}

\makeatletter
\newlength\tmp@\newlength\t@mp
\newcommand{\comp}[3]
  {\mathop{ \settowidth\tmp@{$\displaystyle\mathop{#1}^{#3}_{#2}$}
  \hbox to \tmp@{\hss \settowidth\t@mp{$\displaystyle #1$}\setlength\t@mp{.45\t@mp}
  $\displaystyle\mathop{#1}^{\hspace\t@mp #3}_{\hspace{-\t@mp}#2}$
  \hss} }}
\makeatother
\yil{}
\vol{}
\fpage{}
\lpage{}
\doi{10.3906/elk-2005-155}
\vspace{-3\baselineskip}
\title{On the Outage Performance of SWIPT-NOMA-CRS with imperfect SIC and CSI}
\author[KARA]{
\textbf{Ferdi KARA$^{1}$\thanks{f.kara@beun.edu.tr}}\\
$^{1}$Wireless Communication Technologies Laboratory (WCTLab), Department of Electrical and Electronics, Faculty of Engineering,\\ Zonguldak Bulent Ecevit University, Zonguldak, Turkey, \\ ORCID iD: https://orcid.org/0000-0001-8038-2747\\
\\ [1.8em]
\rec{.202}
\acc{.202}
\finv{..202}
}
\input{elksty.tex}

\begin{document}
\maketitle
\begin{abstract}
In this paper, a non-orthogonal multiple access based cooperative relaying system (NOMA-CRS) is considered to increase spectral efficiency. Besides, the simultaneous wireless information and power transfer (SWIPT) is proposed for the relay in NOMA-CRS. In SWIPT-NOMA-CRS, three different energy harvesting (EH) protocols, power sharing (PS), time sharing (TS) and ideal protocols are implemented. The outage performances of the SWIPT-NOMA-CRS are studied for all three EH protocols. In the analysis, to represent practical/reasonable scenarios, imperfect successive interference canceler (SIC) and imperfect channel state information (CSI) are taken into consideration. The derived outage probability (OP) expressions are validated via computer simulations. Besides, the OP for the benchmark scheme, NOMA-CRS without EH, is also derived under imperfect SIC and CSI. Based on extensive simulations, it is revealed that the SWIPT-NOMA-CRS outperforms NOMA-CRS without EH. Finally, the effects of all parameters on the outage performance of the SWIPT-NOMA-CRS are discussed and for the given scenarios, the optimum PS factor, TS factor and power allocation coefficients are demonstrated.
\end{abstract}
\section{Introduction}
The new era of the wireless communications is beyond the personal communication and it has now many different applications such as Internet of Things (IoT) networks, vehicular communication etc. Hence, the future wireless networks (5G and beyond) are to meet challenging requirements such as very high spectral efficiency, ultra wide coverage and low energy consumption \cite{survey6g}. To this end, the future networks will have the interplay between physical layer techniques such as non-orthogonal multiple access (NOMA) \cite{noma_survey}, cooperative relaying system (CRS) \cite{coop_magazine} and wireless power transfer \cite{swipt_magazine}.

The main idea of the NOMA is to allow multiple users sharing the same resource block. NOMA has generally divided into two groups as code-domain (CD)-NOMA and power-domain (PD)-NOMA. In CD-NOMA, the users share the same resource block with sparse codes, so it is mostly called sparse code multiple access (SCMA) whereas in PD-NOMA, the users are multiplexed with different power allocation (PA) coefficients. Compared to orthogonal multiple access schemes, the main advantage of the NOMA schemes is the spectral efficiency since all users are allocated in the same resource block whereas the main disadvantage is the error performance decay due to the inter-user-interference (IUI) \cite{noma_survey}. Thanks to its potential, NOMA\footnote{This paper deals with PD-NOMA, thus NOMA is used for PD-NOMA after this point.} has been implemented in other physical layer techniques such as cooperative communication \cite{coop2}, visible light communication \cite{vlc1} and index modulation \cite{index1}.

CRS schemes have been studied for two decades because they provide a remarkable performance gain in device-to-device communications. However, due to the cooperative phase, the spectral efficiency of the CRS schemes decreases. To alleviate this performance loss, NOMA-based CRS is proposed in \cite{crs1} where the source implements NOMA in the first phase to increase spectral efficiency and it is proved that NOMA-CRS is superior to conventional CRS. Therefore, NOMA-CRS has attracted recent attention from both academia and industry \cite{crs1,crs2,crs3,crs4,crs5,crs6,crs7,crs8,crs9}. In those works, NOMA-CRS has been investigated in terms of capacity, outage probability (OP) and error probability. Ergodic capacity of the NOMA-CRS is analyzed over Rayleigh \cite{crs1} and Rician \cite{crs2} fading channels. Then, two different NOMA-CRS schemes have been considered in \cite{crs3} and OP expression is derived over Rayleigh fading channels. The error performance of the NOMA-CRS is investigated in \cite{crs4} and a machine learning aided power optimization is proposed to minimize bit error rate. Then, two more NOMA-CRS schemes are proposed in \cite{crs5,crs6}, and capacity/outage performances are investigated. Moreover, NOMA-CRS schemes have been considered when multiple relays are available. NOMA-based diamond relaying, as a subset of the NOMA-CRS with two relays, is analyzed in \cite{crs7} and \cite{crs8} in terms of capacity and error performances, respectively. Contrary to \cite{crs1,crs2,crs3,crs4,crs5,crs6,crs7,crs8}, NOMA-CRS is investigated in \cite{crs9} when an amplify-forward relay is used rather than a decode-forward relay. However, the aforementioned studies assume that the relay node has its own independent (infinite) energy source to help the device-to-device communication. It is neither feasible nor fair for the relay which consumes its energy/battery for a communication where its own symbol is not transmitted. Furthermore, those studies mostly assume perfect successive interference canceler (SIC) at the relay (except for \cite{crs4,crs8}) and perfect channel state information (CSI) at all nodes (except for \cite{crs3}). These are also not practical assumptions and should be relaxed.

On the other hand, simultaneous wireless information and power transfer (SWIPT) is very promising to increase energy efficiency \cite{swipt_magazine}. SWIPT is proved to be practical and it can support nodes with limited energy \cite{swipt1}. To this end, SWIPT has attracted tremendous attention \cite{swipt2,swipt3}. Besides, since it is easily applicable, SWIPT integration into all other physical layer techniques has also taken remarkable consideration \cite{swipt4,swipt5,swipt6,swipt7,swipt8,swipt9,swipt10,swipt11,swipt_noma,kara_swipt}. Specifically, the relay node can harvest its energy to re-transmit the symbols of the source, thus SWIPT-cooperative NOMA schemes have been analyzed in terms of capacity, outage and error performances \cite{swipt4,swipt5,swipt6,swipt7}. SWIPT usage with NOMA has also been investigated in many studies. However, to the best of the author's knowledge, the SWIPT integration into NOMA-CRS has not been studied well although other NOMA-involved systems have been analyzed with SWIPT \cite{swipt8,swipt9,swipt10,swipt11}. Only two studies in the literature consider NOMA-CRS with SWIPT. In \cite{swipt_noma}, the authors consider a CRS where the source transmits two symbols in two time slots and in the second phase, an energy harvesting relay re-transmits the symbol of the first phase. The considered model in \cite{swipt_noma} is different from the conventional NOMA-CRS schemes \cite{crs1,crs2,crs3,crs4,crs5,crs6,crs7,crs8,crs9} since a power allocation is not implemented. Nevertheless, it is called as SWIPT-CRS deploying NOMA since an SIC should be performed at the destination due to the second phase. The authors provide outage probability analysis for the considered model. Then, in \cite{kara_swipt}, the author's preliminary work, a NOMA-CRS with SWIPT is considered and only the achievable rate is analyzed. However, these two studies assume perfect SIC at the relay/destination and perfect CSI at all nodes. Thus, these assumptions should be relaxed and the analysis should be further extended for practical scenarios.

Based on above discussions, in this paper, the NOMA-CRS with SWIPT is proposed where the relay harvests its energy from radio frequency (RF) waves to transmit symbols. In SWIPT integration, three different energy harvesting (EH) protocols, power sharing (PS), time sharing (TS) and ideal protocols, are implemented. All non-practical assumptions are relaxed, to this end, the OP expressions for all EH protocols are derived with imperfect SIC and CSI. Besides, the OP for the benchmark, NOMA-CRS without EH, is also derived. It is revealed that the SWIPT-NOMA-CRS outperforms NOMA-CRS without EH. Then, the effects of EH factors and PA coefficient are discussed for the outage performance of the SWIPT-NOMA-CRS.

The rest of the paper is organized as follows. In Section 2, the SWIPT-NOMA-CRS is introduced where all EH protocols and power transfers are defined. The information transferring in SWIPT-NOMA-CRS and the related signal-to-interference plus noise (SINR) definitions are also given in this section. Then, in Section 3, the theoretical analysis of OP is performed. In Section 4, the analysis is validated via computer simulations and performance comparisons are presented. Finally, Section 5 discusses the results and concludes the paper.
\section{System Model}
A NOMA-based relaying system is considered for a device-to-device communication where a source (S), a decode-forward (DF) relay (R) and a destination (D) are located. All nodes are assumed to be equipped with single antenna. The channel fading coefficient between each node follows $CN(0,\Omega_k), k=\{sr, sd, rd\}$ where $\Omega_k$ denotes the large-scale path loss component. The imperfect channel state information (CSI) is considered and the estimated channel at each node is given by $\hat{h}_k=h_k+\epsilon$ where $\epsilon= CN(0,\kappa)$ which is an appropriate model for the practical channel estimation techniques. The DF relay operates in half duplex mode. Thus, the total communication is completed in two phases. In order to alleviate inefficiency of the CRS, NOMA is implemented at the source so that the spectral efficiency is increased. Besides, it is assumed that the relay do not have independent energy source and it harvests its energy from the RF signal in the first phase to transmit symbols in the second phase. Therefore, the source implements a SWIPT so that the system is called SWIPT-NOMA-CRS. The illustration of the SWIPT-NOMA-CRS is given in Figure 1.
\begin{figure}[!t]
   \centering
     \includegraphics[width=14cm]{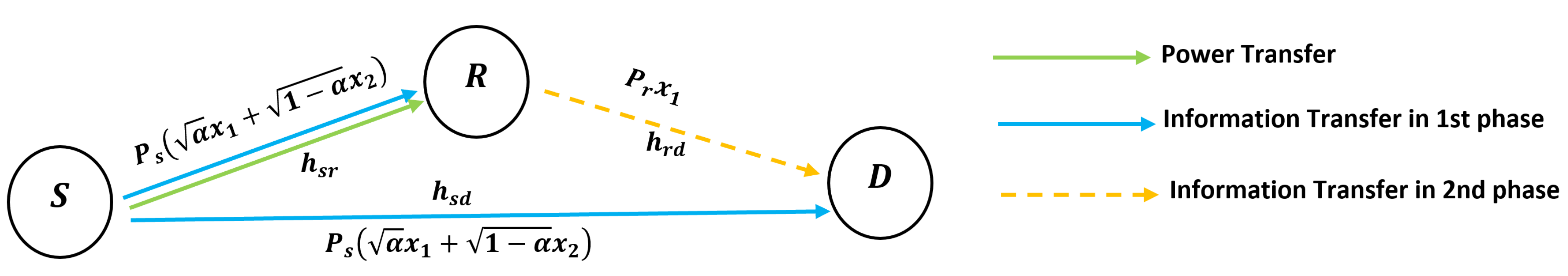}
    \caption{The illustration of SWIPT-NOMA-CRS }
    \label{fig1}
 \end{figure}
\subsection{Transmit Power and Energy Harvesting}
In the energy-constrained networks, a node can harvest energy from the RF signals so that it can use it to transmit symbols. In the EH, the RF signals are converted to Direct Current (DC) power via energy receiver (called rectenna) \cite{swipt2}. This rectenna can be placed on the same circuit with a transceiver, hence the SWIPT becomes possible. According to internal characteristics (due to the diode ) of the rectenna, the EH protocols are categorized in two groups : 1) linear EH 2) non-linear EH. In this paper, the linear EH protocols are considered where the received RF waves are converted a DC power with a energy conversion coefficient.

As explained above, the relay node harvests its energy from the RF signal transmitted by the source. In this paper, within the linear EH models, three different EH protocols (i.e., PS, TS and ideal protocols) \cite{swipt7} are implemented. Time schedules for all three EH protocols are shown in Figure2. In Figure2, the time schedule of the benchmark is also given where no EH is implemented and the total energy is shared among the source and relay.
\subsubsection{Benchmark (No EH)}
As considered in the literature, if the relay has no ability to harvest energy, it has its own energy source, and the total consumed energy/power is the sum of the energy/power consumed by the source and the relay. As shown in Figure 2(a), the source and the relay consume their energy within $\sfrac{T}{2}$ seconds. Thus, for fairness, considering the total energy consumption during $T$ seconds, the transmit powers of both source and relay are given as
\begin{equation}
 P_s=P_r=P_T.
\end{equation}
where $P_s$ and and $P_r$ are the transmit powers of the source and relay. $P_T$ is the total consumed power during the whole communication ($T$ seconds).
\subsubsection{Power Sharing (PS) Protocol}
As shown in Figure 2(b), in the PS protocol, the communication times from source-to-relay (S-R) and from relay-to-destination (R-D) are equal and cover half of the total $T$ duration. Thus, by considering the total consumed energy during $T$ seconds, the transmit power of the source within $\sfrac{T}{2}$ seconds is equal to
\begin{equation}
 P_s=2P_T.
\end{equation}
The relay harvests its energy during the first $\sfrac{T}{2}$ seconds, hence the harvested energy in PS protocol is given as
 \begin{equation}
       E_H=\eta\rho P_s \left|\hat{h}_{sr}\right|^2 (\sfrac{T}{2}),
 \end{equation}
where $\eta$ is the energy conversion coefficient and it is given as $0<\eta<1$. In (3), $\rho$ is the PS factor as represented in Figure 2(b). It is noteworthy that the energy is harvested according to the estimated channel $\hat{h}_{sr}$. The harvested energy is consumed by the relay to transmit symbols between R-D within the remained $\sfrac{T}{2}$ seconds. Therefore, the transmit power of the relay in PS protocol is obtained as
 \begin{equation}
       P_r=\eta\rho P_s \left|\hat{h}_{sr}\right|^2.
 \end{equation}
 \begin{figure}[!t]
   \centering
     \includegraphics[width=17cm]{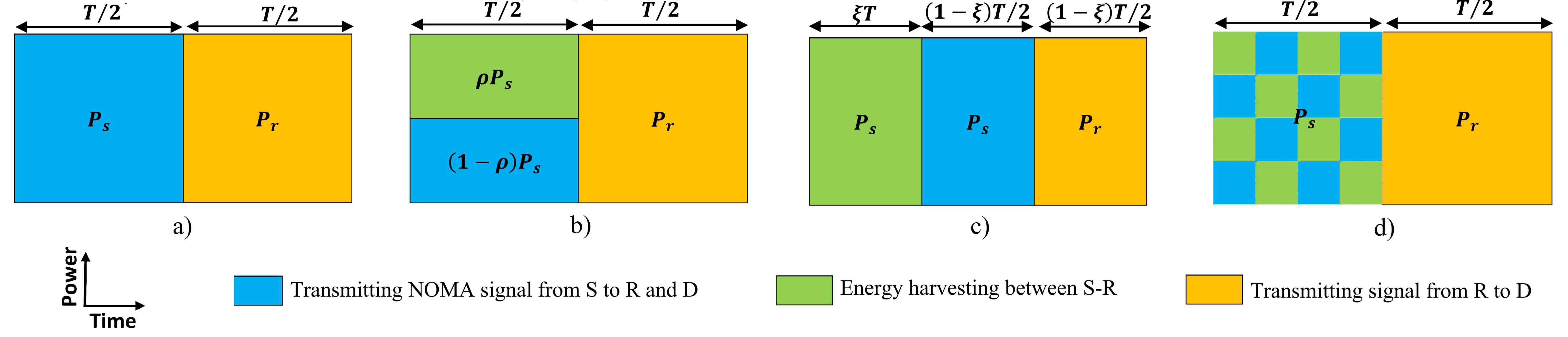}
    \caption{Time schedules for a) Benchmark (No EH) b) EH with PS protocol c) EH with TS protocol d) EH with ideal protocol }
    \label{fig2}
 \end{figure}
\subsubsection{Time Sharing (TS) Protocol}
As shown in Figure 2(c), in TS protocol, the source transmits power for EH in the first $\xi T$ seconds and then, it transmits information (data) in the next $\sfrac{(1-\xi)T}{2}$ second. Thus, the total consumed energy by the source in TS mode is given by  $\sfrac{P_s(1+\xi)T}{2}$. For fairness, regarding the total consumed energy in $T$ seconds with total power $P_T$, the source power in TS protocol is obtained as
\begin{equation}
   P_s=\frac{2P_T}{(1+\xi)},
\end{equation}
where $\xi$ is the TS factor as given in Figure 2(c). The harvested energy from this source power within $\xi T$ seconds is given as
 \begin{equation}
       E_H=\eta P_s\hat{h}_{sr} (\xi T).
 \end{equation}
Since this harvested energy at the relay is consumed within the remained $\sfrac{(1-\xi)}{2}$ seconds, the transmit power of the relay in TS protocol is derived as
\begin{equation}
      P_r=2\eta\frac{\xi}{(1-\xi)} P_s \hat{h}_{sr}.
 \end{equation}
\subsubsection{Ideal Protocol}
In the ideal protocol, as shown in Figure 2(d), the source transmits power during $\sfrac{T}{2}$ seconds. Thus, again by considering the total consumed power within $T$ seconds, the transmit power of the source, in ideal protocol, is given as
\begin{equation}
 P_s=2P_T.
\end{equation}
During the first $\sfrac{T}{2}$ seconds, the relay harvests energy from the RF wave between S-R, thus the harvested energy is given as
\begin{equation}
       E_H=\eta P_s \left|\hat{h}_{sr}\right|^2 (\sfrac{T}{2}).
 \end{equation}

It is noteworthy that the energy and data transfers are achieved with the same power $P_s$, in the ideal protocol whereas the transmit power of the source is allocated by $\rho$ for energy and data transfers in the PS protocol. This is the main difference between the PS and ideal protocols.

Then, this harvested energy is consumed by the relay to transmit symbols between R-D link within $\sfrac{T}{2}$ seconds. Therefore, the transmit power of the relay, in ideal protocol, is obtained as
 \begin{equation}
       P_r=\eta P_s \left|\hat{h}_{sr}\right|^2.
 \end{equation}
\subsection{Information Transfer}
As explained above and shown in Figure 1, the information transfer is completed in two phases. In the first phase, the source implements NOMA for the consecutive two symbols of the destination to increase spectral efficiency and broadcasts this superposition coded symbol to both relay and destination. Thus, the received signals by both nodes are given as
\begin{equation}
 y_k=\sqrt{pP_s}\left(\sqrt{\alpha}x_1+\sqrt{\left(1-\alpha\right)}x_2\right)h_k+n_k, \ k=sr,sd,
\end{equation}
where $\alpha$ is the PA coefficient (i.e., $\alpha<0.5$) among the symbols. $n_k$ is the additive white Gaussian noise (AWGN) and follows $CN(0,\sigma^2)$. It is hereby noted that the $h_k$ in (11) is the actual channel fading coefficient between nodes and it is estimated as $\hat{h}_k=h_k+\epsilon$ at the receiving nodes as explained above. Besides, it is noteworthy that the transmit power of the source changes according to the used EH protocol as explained in detail in the previous subsection (see eq. (1), (4), (7) and (10)). Thus, the $p$ coefficient denotes the ratio how much the source power is allocated for the information transfer. According to the EH protocol, it is given as
\begin{equation}
    p=\begin{cases}
    1-\rho, & \text{in PS protocol}, \\
    1, & \text{in TS, ideal protocols and in no EH}.
    \end{cases}
\end{equation}
In the first phase, according to the received signal $y_{sd}$, $x_2$ symbols are detected with a conventional detector (e.g., maximum likelihood (ML)) at the destination by pretending $x_1$ symbols as a noise. At the same time, the relay implements a successive interference canceler (SIC) to detect $x_1$ symbols. In the SIC process, the relay firstly detects $x_2$ symbols with a ML detector, then it subtracts these estimated $x_2$ symbols from the received signal $y_{sd}$. Finally, it implements one more ML detector to detect $x_1$ symbols based on the remained signal after subtraction. In the second phase, the relay re-transmits the detected $x_1$ symbols to the destination. Hence, the received signal in the second phase is given as
\begin{equation}
        y_{rd}=\sqrt{P_r}\hat{x}_1h_{rd}+n_{rd},
 \end{equation}
 where $\hat{x}_1$ is the detected symbol at the relay after SIC. It is worth noting that
 $P_r$ is the transmit power of the relay and changes according to the EH protocol since it is harvested from the source-relay link in the first phase (or it has its own energy in no EH benchmark) (see eq. (1), (4), (7) and (10)). Besides, the channel fading coefficient $h_{rd}$ is estimated as $\hat{h}_{rd}$ at the destination. Lastly, the destination detects $x_1$ symbols based on $y_{rd}$.
\subsection{Received Signal-to-Interference plus Noise Ratios (SINRs)}
As given in (11), the source transmits superposition-coded NOMA symbols in the first phase, thus an interference occurs. Both relay and destination detect $x_2$ symbols firstly by pretending $x_1$ symbols as noise. Therefore, by considering imperfect CSI at the nodes, the received SINRs for the $x_2$ symbols at the nodes are given as
\begin{equation}
\begin{split}
    &SINR_{x_2}^{(sr)}=\frac{(1-\alpha)pP_s \hat{\gamma}_{sr}}{\alpha pP_s\hat{\gamma}_{sr}+pP_s\kappa+\sigma^2}, \\
     &SINR_{x_2}^{(sd)}=\frac{(1-\alpha)pP_s\hat{\gamma}_{sd}}{\alpha pP_s\hat{\gamma}_{sd}+pP_s\kappa+\sigma^2}.
\end{split}
\end{equation}
where $\hat{\gamma}_{k}\triangleq|\hat{h}_k|^2$ is defined and it follows exponential distribution with the parameter $\hat{\Omega}_k=\Omega_k-\kappa$. In (14), the first terms in the denominators define the interference due to the $x_1$ symbols (NOMA signalling) whereas the second and the third terms are the effects of the imperfect CSI and the AWGN, respectively.

On the other hand, the relay implements SIC to detect $x_1$ symbols in the first phase, thus, by also considering imperfect SIC, the received SINR for the $x_1$ symbols at the relay is given as
\begin{equation}
    SINR_{x_1}^{(sr)}=\frac{\alpha pP_s \hat{\gamma}_{sr}}{(1-\alpha) pP_s|g|^2+pP_s\kappa+\sigma^2},
\end{equation}
where the first term in the denominator denotes the imperfect SIC effect where $g$ follows $CN(0,\delta\Omega_{sr})$. $0<\delta<1$ is defined where $\delta=0$ and $\delta=1$ denote perfect SIC and no SIC cases, respectively. Just like (14), in (15), the second and the third terms of the denominator are the effects of the imperfect CSI and the AWGN, respectively. In (14) and (15), the transmit power of the source is given in (1), (2), (5) and (8) according to the EH protocol.

Lastly, the received SINR in the second phase is given as
\begin{equation}
    SINR_{\hat{x}_1}^{(rd)}=\frac{P_r \hat{\gamma}_{rd}}{ P_r\kappa+\sigma^2}.
\end{equation}
It is important to note that $P_r$ in (16) will include the $\hat{\gamma}_{sr}$ random variable if any of EH protocols is implemented (see (4), (7) and (10)). It can be given as independent from $\hat{\gamma}_{sr}$ only if no EH is implemented (1).
\section{Outage Probability (OP) Analysis}
The outage event for a communication system is defined as the probability of the achievable rate being below the target rate (QoS). To this end, the outage probability for the symbols in SWIPT-NOMA-CRS is given by
\begin{equation}
    P_i(out)=P(R_i<\acute{R}_i), \ i=1,2,
\end{equation}
where $R_i$ and $\acute{R}_i$ denote the achievable rate and the target rate of  $x_i, \ i=1,2$ symbols. Hence, as firstly, the achievable rates of the symbols should be defined. Since a cooperative communication is considered, the achievable rate of $x_1$ symbols is limited by the weakest link \cite{conv_crs}. Thus, according to the Shannon Theory \cite{shannon}, the achievable rate of $x_1$ symbols is given as
\begin{equation}
    R_1=\zeta B\log_2\left(1+\min\{SINR_{x_1}^{(sr)},SINR_{\hat{x}_1}^{(rd)}\}\right),
\end{equation}
where $B=\sfrac{1}{T}$ the bandwidth. In (18), $\zeta$ exists since the total communication is handled in two phases. According to the time schedules of EH protocols in Figure 2 and Section 2.1, it is defined as
\begin{equation}
    \zeta=\begin{cases}
    \frac{1-\xi}{2}, &\text{in TS protocol} \\
    \frac{1}{2}, &\text{in PS and ideal protocols and in no EH}
    \end{cases}
\end{equation}
On the other hand, although cooperative communication is not considered for the $x_2$ symbols, to guarantee SIC operation, the achievable rate at the relay for $x_2$ symbols should not also cause outage. Thus, the achievable rate of $x_2$ symbols is also given as \cite{crs1}
\begin{equation}
    R_2=\zeta B\log_2\left(1+\min\{SINR_{x_2}^{(sr)},SINR_{x_2}^{(sd)}\}\right).
\end{equation}
Firstly, to obtain $P_2(out)$, by substituting (14) into (20) then into (17), the OP\footnote{In the following OP analysis, $B$ is removed for notation simplicity since it is equal in all scenarios and does not affect the analysis.} is defined as
\begin{equation}
    P_2(out)=P\left(\min\left\{\frac{(1-\alpha)pP_s \hat{\gamma}_{sr}}{\alpha pP_s\hat{\gamma}_{sr}+pP_s\kappa+\sigma^2},\frac{(1-\alpha)pP_s\hat{\gamma}_{sd}}{\alpha pP_s\hat{\gamma}_{sd}+pP_s\kappa+\sigma^2}\right\}<\phi_2\right),
\end{equation}
where $\phi_i= 2^{\sfrac{\acute{R}_i}{\zeta}}-1, \ i=1,2$. The probability of $P(Z<z)$ is represented by $F_Z(z)$ which is called the cumulative distribution function (CDF) of $Z$. If  $Z=\min\{X,Y\}$ is defined and in case X and Y are statistically independent, the CDF of the Z is given by $F_Z(z)=F_X(z)+F_Y(z)-F_X(z)F_Y(z)$ \cite{prob} where $F_X()$ and $F_Y()$ are the CDFs of $X$ and $Y$ random variables, respectively. Recalling $\hat{\gamma}_{sr}$ and $\hat{\gamma}_{sd}$ are statistically independent, $X\triangleq\frac{(1-\alpha)pP_s \hat{\gamma}_{sr}}{\alpha pP_s\hat{\gamma}_{sr}+pP_s\kappa+\sigma^2}$, $Y\triangleq\frac{(1-\alpha)pP_s\hat{\gamma}_{sd}}{\alpha pP_s\hat{\gamma}_{sd}+pP_s\kappa+\sigma^2}$ and/or $\hat{\gamma}_{sr}\triangleq\frac{X\left(pP_s\kappa+\sigma^2\right)}{\left(1-(1+X)\alpha\right)pP_s}$, $\hat{\gamma}_{sd}\triangleq\frac{Y\left(pP_s\kappa+\sigma^2\right)}{\left(1-(1+Y)\alpha\right)pP_s}$ are defined. Therefore, the OP of $x_2$ symbols is given as
\begin{equation}
\begin{split}
   P_2(out)=F_X(\phi_2)+F_Y(\phi_2)-F_X(\phi_2)F_Y(\phi_2)=F_{\hat{\gamma}_{sr}}\left(A_1\right)+F_{\hat{\gamma}_{sd}}\left(A_1\right)-F_{\hat{\gamma}_{sr}}\left(A_1\right)F_{\hat{\gamma}_{sd}}\left(A_1\right),
\end{split}
\end{equation}
where $A_1\triangleq\frac{\phi_2\left(pP_s\kappa+\sigma^2\right)}{\left(1-(1+\phi_2)\alpha\right)pP_s}$ is defined for representation/notation simplicity. As defined in (14), $\hat{\gamma}_{k}=|\hat{h}_k|^2$ follows exponential distribution, hence the CDF of the $\hat{\gamma}_k$ is $F_{\hat{\gamma}_k}(.)=1-\exp\left(-\sfrac{\hat{\gamma}_k}{\hat{\Omega}_k}\right)$ where $\hat{\Omega}_k=\Omega_k-\kappa$ is defined. Thus, the OP of the $x_1$ symbols is derived as
\begin{equation}
    P_2(out)=\left\{1-\exp\left(-\frac{A_1}{\hat{\Omega}_{sr}}\right)\right\}+\left\{1-\exp\left(-\frac{A_1}{\hat{\Omega}_{sd}}\right)\right\}-\left\{1-\exp\left(-\frac{A_1}{\hat{\Omega}_{sr}}\right)\right\}\left\{1-\exp\left(-\frac{A_1}{\hat{\Omega}_{sd}}\right)\right\}.
\end{equation}
Likewise in the analysis of $x_2$ symbols, the OP of the $x_1$ symbols is obtained, by substituting (15) and (16) into (18) then into (17), as
\begin{equation}
    P_1(out)=P\left(\min\left\{\frac{\alpha pP_s \hat{\gamma}_{sr}}{(1-\alpha) pP_s|g|^2+pP_s\kappa+\sigma^2},\frac{P_r \hat{\gamma}_{rd}}{ P_r\kappa+\sigma^2}\right\}<\phi_1\right).
\end{equation}
Recalling $P_r$ is a function of $P_s$ and $\bar{\gamma}_{sr}$ based on the EH, the OP of the $x_1$ symbols is rewritten as
\begin{equation}
    P_1(out)=P\left(\min\left\{\frac{\alpha pP_s \hat{\gamma}_{sr}}{(1-\alpha) pP_s|g|^2+pP_s\kappa+\sigma^2},\frac{\Upsilon P_s \hat{\gamma}_{sr}\hat{\gamma}_{rd}}{ \Upsilon P_s \hat{\gamma}_{sr} \kappa+\sigma^2}\right\}<\phi_1\right)
\end{equation}
where $\Upsilon$ is the power transformation coefficient from the power of the source in that EH protocol and according to section 2.1, it is given as
\begin{equation}
\Upsilon=\begin{cases}
\eta\rho, & \text{in PS protocol}, \\
2\eta\frac{\xi}{1-\xi}, & \text{in TS protocol}, \\
\eta, & \text{in ideal protocol}. \\
\end{cases}
\end{equation}
In (25), since both SINRs include $\hat{\gamma}_{sr}$ they are statistically correlated. Besides, the second SINR includes the same random variable on both nominator and denominator, thus the joint CDF is very hard to obtain. Nevertheless, without loss of generality, to derive a very tight approximate expression, the rule of independent random variables given in (22) is applied. The OP of the $x_1$ symbols is obtained as
\begin{equation}
    P_1(out)\cong F_{\hat{\gamma}_{sr}}\left(A_2\right)+F_{\hat{\gamma}_{sr},\hat{\gamma}_{rd}}\left(\phi_1\right)-F_{\hat{\gamma}_{sr}}\left(A_2\right)F_{\hat{\gamma}_{sr},\hat{\gamma}_{rd}}\left(\phi_1\right),
\end{equation}
where $A_2\triangleq\frac{\phi_1\left((1-\alpha) pP_s\delta\hat{\Omega}_{sr}+pP_s\kappa+\sigma^2\right)}{\alpha pP_s}$ is defined for notation simplicity. $F_{\hat{\gamma}_{sr},\hat{\gamma}_{rd}}$ is the joint CDF for the second SINR term in (25). Since the $\hat{\gamma}_{sr}$ is exponentially distributed, the first term in (27) can be easily obtained as
\begin{equation}
   F_{\hat{\gamma}_{sr}}\left(A_2\right)=1-\exp\left(-\frac{A_2}{\hat{\Omega}_{sr}}\right).
\end{equation}
The second term in (27) is obtained as
\begin{equation}
    \begin{split}
     F_{\hat{\gamma}_{sr},\hat{\gamma}_{rd}}\left(\phi_1\right)&=\comp{\iint}{0}{\phi_1}\frac{\Upsilon P_s \hat{\gamma}_{sr}\hat{\gamma}_{rd}}{ \Upsilon P_s \hat{\gamma}_{sr} \kappa+\sigma^2}d\hat{\gamma}_{sr}d\hat{\gamma}_{rd},
     =\int\limits_0^\infty\frac{1}{\hat{\Omega}_{rd}}\exp\left(-\sfrac{\hat{\gamma}_{rd}}{\hat{\Omega}_{rd}}\right)\int\limits_{\frac{\phi_1\sigma^2}{\left(\hat{\gamma}_{rd}-\phi_1\kappa\right)\Upsilon P_s}}^\infty\frac{1}{\hat{\Omega}_{sr}}\exp\left(-\sfrac{\hat{\gamma}_{sr}}{\hat{\Omega}_{sr}}\right)d\hat{\gamma}_{sr}d\hat{\gamma}_{rd}, \\
     &=1-\int\limits_0^\infty\frac{1}{\hat{\Omega}_{rd}}\exp\left(-\frac{\hat{\gamma}_{rd}}{\hat{\Omega}_{rd}}-\frac{\phi_1\sigma^2}{\left(\hat{\gamma}_{rd}-\phi_1\kappa\right)\Upsilon P_s\hat{\Omega}_{sr}}\right)d\hat{\gamma}_{rd}.
    \end{split}
\end{equation}
Since the exponential expression includes a complex polynomial in (29), to the best of the author's knowledge, it has no closed-form solution. Nevertheless, it can be easily computed by numerical tools such as MATLAB, MAPPLE, MATHEMATICA. By substituting (28) and (29), the OP of $x_1$ symbols is derived.

Since both symbols are transmitted to the same destination in NOMA-CRS, the system is considered to be in outage in case any of the symbols being in outage. Therefore, the OP of the SWIPT-NOMA-CRS is equal to the union outage probabilities of the symbols and it is given by
\begin{equation}
    P_{SWIPT-NOMA-CRS}(out)=P_1(out)\cup P_2(out)=P_1(out)+P_2(out)-P_1(out)P_2(out).
\end{equation}
It is derived by substituting (23) and (27) into (30).
\subsection{Benchmark Analysis (NOMA-CRS without EH)}
As explained in the system model, the first phases in SWIPT-NOMA-CRS and NOMA-CRS without EH are the same. The only difference is the power of the source. Therefore, the OP of $x_2$ symbols in NOMA-CRS without EH is also equal to (23).

On the other hand, since no EH is implemented, the relay will have its own energy thereby the SINR between R-D will be independent from the first phase (do not include $\hat{\gamma}_{sr}$ anymore). Therefore, by using (24) and considering that  $\hat{\gamma}_{sr}$ and $\hat{\gamma}_{rd}$ are statistically independent, the OP of $x_1$ symbols is obtained, by repeating steps between (21)-(23), as
\begin{equation}
    P_1(out)=\left\{1-\exp\left(-\frac{A_2}{\hat{\Omega}_{sr}}\right)\right\}+\left\{1-\exp\left(-\frac{A_3}{\hat{\Omega}_{rd}}\right)\right\}-\left\{1-\exp\left(-\frac{A_2}{\hat{\Omega}_{sr}}\right)\right\}\left\{1-\exp\left(-\frac{A_3}{\hat{\Omega}_{rd}}\right)\right\}
\end{equation}
where $A_3\triangleq\frac{\phi_1\left(P_r\kappa+\sigma^2\right)}{P_r}$.
Lastly, the overall OP of the NOMA-CRS without EH is derived by substituting (23) and (31) into (30).
\section{Numerical Results}
In this section, the derived OP expressions of the SWIPT-NOMA-CRS are validated for all EH protocol and without EH (benchmark).
In the following figures, the lines denote the theoretical curves whereas the markers represent computer simulations. Besides, unless the figures are presented with respect to one of the parameters, the simulation parameters are given in Table. Moreover, in the figures, the curves with the same parameters (e.g., $\kappa$, $\alpha$) are noted with dashed circles.

\begin{table}[!t]
\centering
\caption{Simulation parameters}
\begin{tabular}{c|c||c|c}
\hline
\textbf{Parameter} &\textbf{Value} & \textbf{Parameter} &\textbf{Value} \\ \hline
$\Omega_k, \ k=sr,sd,rd$ & $ [10,2,10] $ &Time sharing factor  ($\xi$) & $0.2$ \\ \hline
Imperfect CSI factor ($\kappa$)   & $0$ (perfect CSI), $0.01$&Imperfect SIC factor ($\delta$)   & $0$ (perfect SIC), $0.001$\\ \hline
Power allocation ($\alpha$) & $0.1$ and $0.2$ &Bandwidth ($B$) & 1 MHz\\ \hline
Energy conversation coefficient ($\eta$) & $0.95$ &Target Rates ($\acute{R}_i, \ i=1,2$) & $[500Kbps,100Kbps]$ \\ \hline
Power sharing factor ($\rho$) & $0.2$ & & \\ \hline
\end{tabular}
\label{table1}
\end{table}

In Figure 3 and Figure 4, the outage performances of $x_1$ symbols are presented for perfect SIC ($\delta=0$) and imperfect SIC ($\delta=0.001$) cases, respectively. In both figures, the results are given for two different PA coefficient and for both perfect CSI and imperfect CSI cases. Firstly, it is noteworthy that the derived OP expressions match well with simulations for all EH protocols which proves that the analysis is very tight. Besides, the derived OP for without EH protocol is perfectly-matched with simulations. In both figures, the OP of $x_1$ symbols have better performances in all SWIPT-NOMA-CRS schemes (the EH protocols are implemented) rather than NOMA-CRS without EH (benchmark). For instance, according to the EH protocol, the SWIPT-NOMA-CRS schemes achieve the same outage performance with $\sim 2.5\ dB$ to $4\ dB$ less power than NOMA-CRS without EH. This is very promising for energy efficiency. However, with the increase of SIC uncertainties, the advantage of the TS protocol can be diminished in high SNR region. This is explained as follows. The transmit power of the source is higher in SWIPT-NOMA-CRS (with EH protocols). Therefore, the effect of imperfect SIC becomes greater in the first phase in SWIPT-NOMA-CRS since the imperfect SIC effect depends on the source power given in (15). In addition to this, in TS protocol, since the power transfer and information transfer are achieved by time division duplex, the information transfer is implemented with a lower duration and this increases $\phi_i$ in TS protocol due to the $\zeta$ (see (19) and below (21)). However, one can easily see that the probability of the imperfect SIC becomes very low in SWIPT-NOMA-CRS when an actual modulator/detector is implemented since, with the increase of the transmit power, the error probability decreases significantly. Then, in Figure 5, the outage performance of $x_2$ symbols are presented for two different PA and CSI conditions. As seen from the figures, the derived OP expressions match perfectly with simulations for all EH and without EH protocols. The SWIPT-NOMA-CRS is again superior to NOMA-CRS in terms of outage performances of $x_2$ symbols in all scenarios. Indeed, this performance gain is more than the one of $x_1$ symbols. The SWIPT-NOMA-CRS can provide $\sim 3\ dB$ to $5\ dB$ energy efficiency.
 \begin{figure}[!t]
   \centering
     \includegraphics[width=16cm]{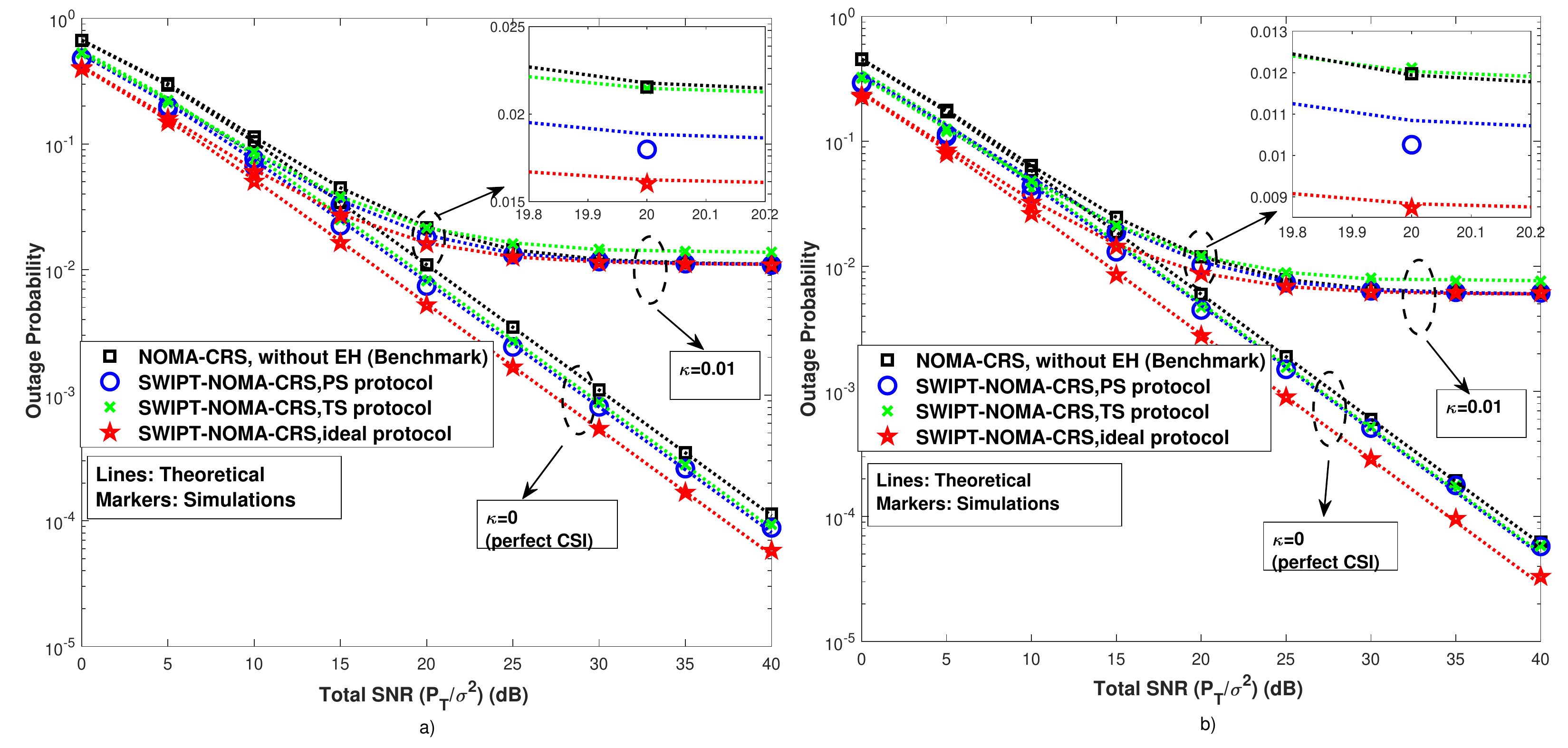}
    \caption{Outage Performance of $x_1$ symbols with perfect SIC ($\delta=0$) a) PA $\alpha=0.1$ b) PA $\alpha=0.2$.}
    \label{fig3}
 \end{figure}
   \begin{figure}[!t]
   \centering
     \includegraphics[width=16cm]{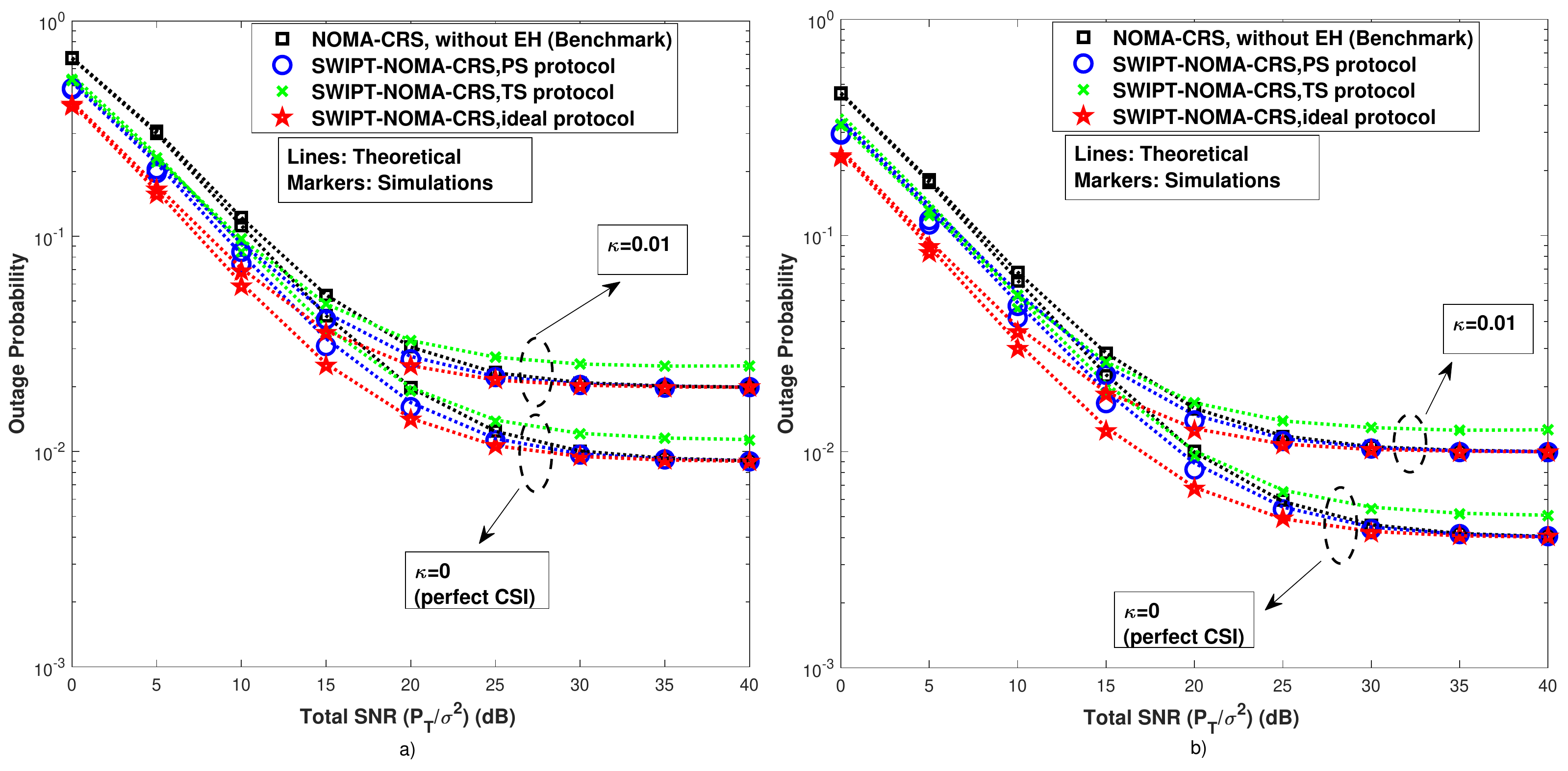}
    \caption{Outage Performance of $x_1$ symbols with imperfect SIC ($\delta=0.001$)a) PA $\alpha=0.1$ b) PA $\alpha=0.2$.}
    \label{fig4}
 \end{figure}
   \begin{figure}[!t]
   \centering
     \includegraphics[width=16cm]{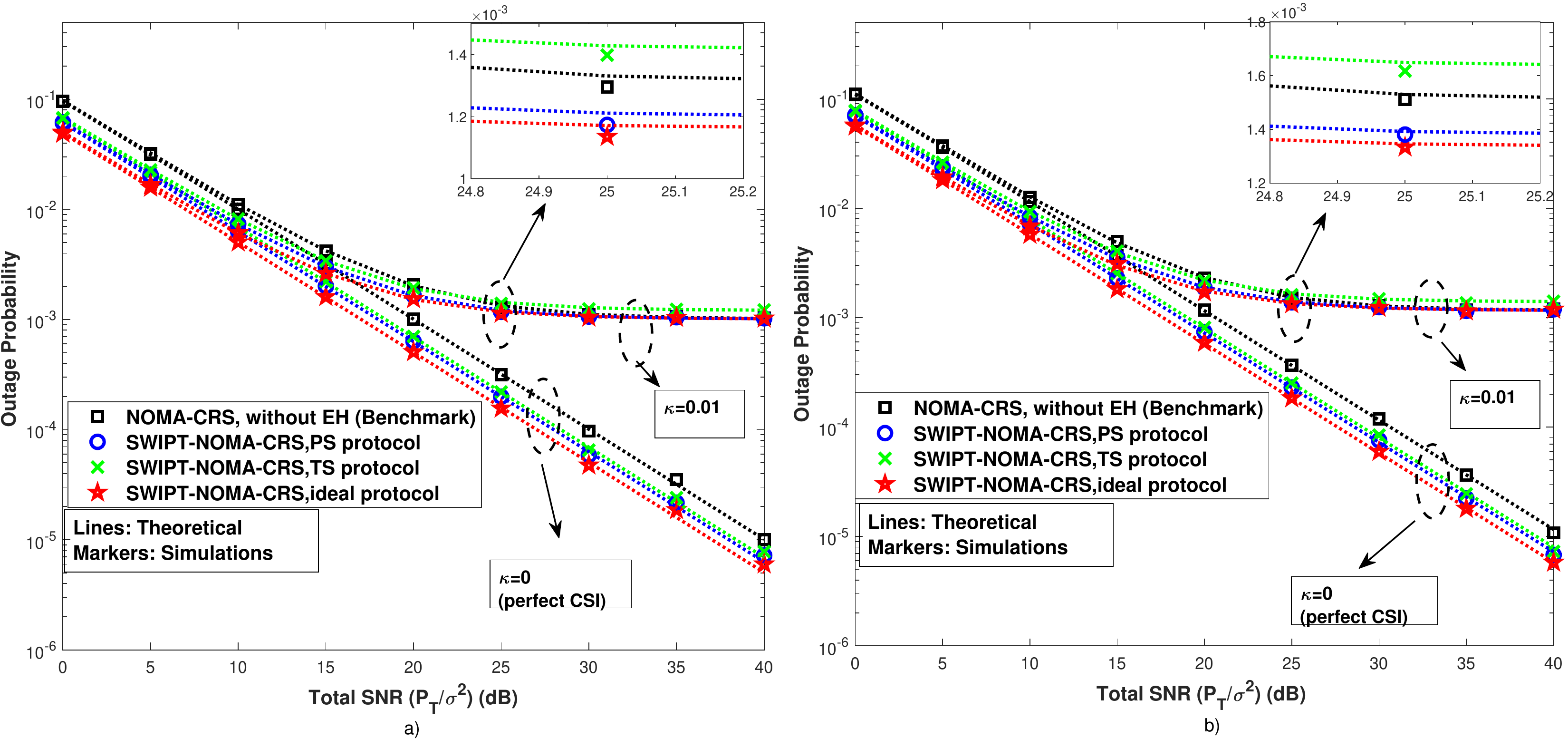}
    \caption{Outage Performance of $x_2$ symbols a) PA $\alpha=0.1$ b) PA $\alpha=0.2$.}
    \label{fig5}
 \end{figure}

Since the validations of the derived OP expressions for both symbols are presented in Figure 3-Figure 5 and the overall outage probability of the SWIPT-NOMA-CRS is defined as the union of the OPs of the symbols (30), in the following figures, only the OP of the SWIPT-NOMA-CRS (union OP of the symbols) is presented. In order to investigate the effect of the imperfect SIC, in Figure 6, the OP of the SWIPT-NOMA-CRS is given with respect to imperfect SIC effect ($\delta$). In Figure 6, the total transmit SNR is assumed to be $30dB$. As seen in the previous figures, with the increase of the imperfect SIC effect, the gap between performances becomes lower. Nevertheless, as expected from the previous results, the PS and ideal protocols in SWIPT-NOMA-CRS still outperforms NOMA-CRS without EH even a bit whereas the TS protocol performs worse with higher imperfect SIC effects. Besides, if the imperfect SIC effect becomes too high (e.g. $\delta\geq -5\ dB$), all considered scenarios are always in outage. In Figure 6(a) and Figure 6(b), one can see that the imperfect CSI does not have too much effect on the outage performance when imperfect SIC is higher, since the performance is dominated by the imperfect SIC. This is also seen in comparisons of PA coefficients. Since the imperfect SIC causes worse performance of $x_1$ symbols and it pulls down the overall performance of SWIPT-NOMA-CRS, the better performance is achieved by increasing the allocated power to $x_1$ symbols (e.g., $\alpha=0.2$).
 \begin{figure}[!t]
   \centering
     \includegraphics[width=16cm]{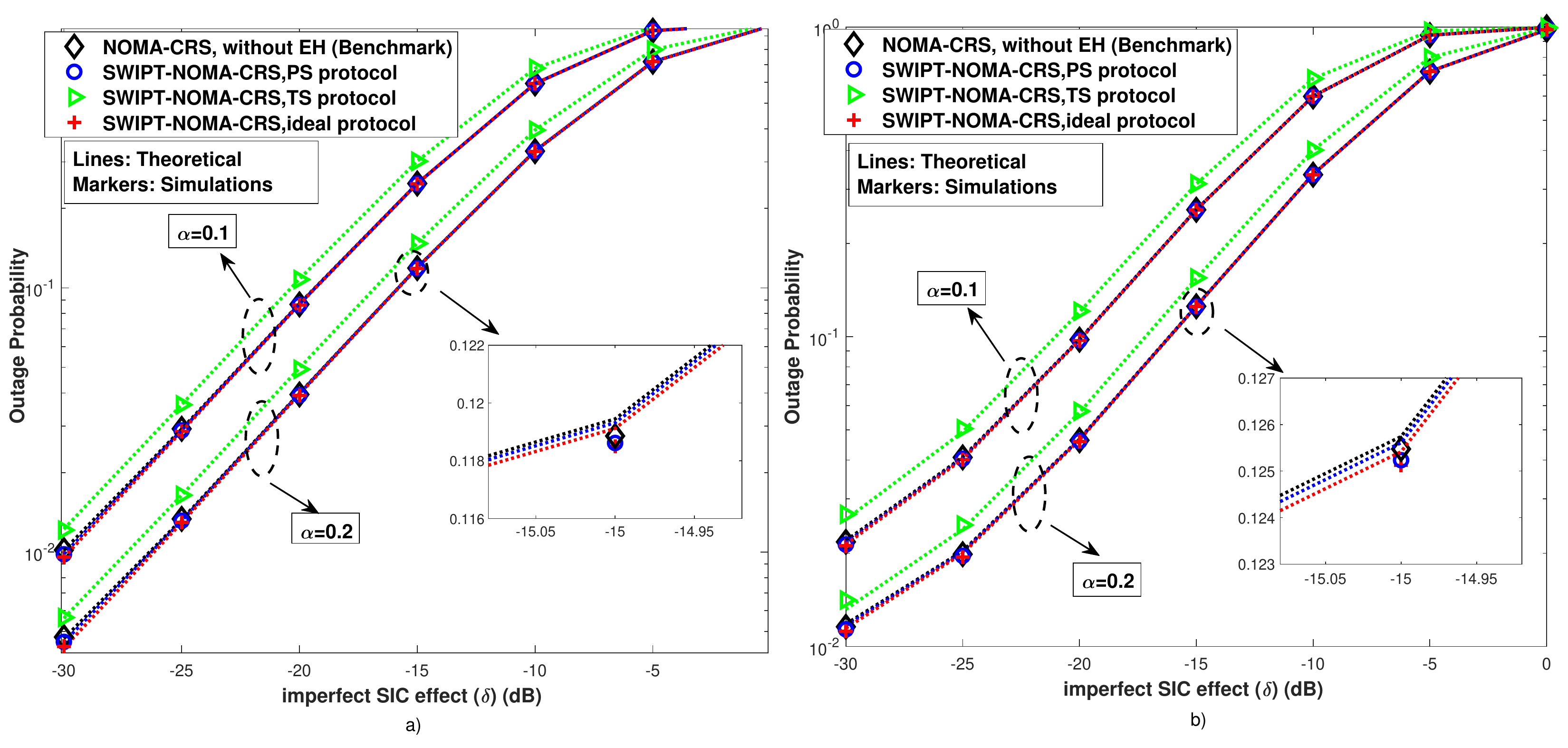}
    \caption{Outage Performance of the SWIPT-NOMA-CRS with respect to imperfect SIC effect ($\delta$) a) $\kappa=0$ (perfect CSI) b) $\kappa=0.01$ (imperfect CSI).}
    \label{fig6}
 \end{figure}
   \begin{figure}[!t]
   \centering
     \includegraphics[width=16cm]{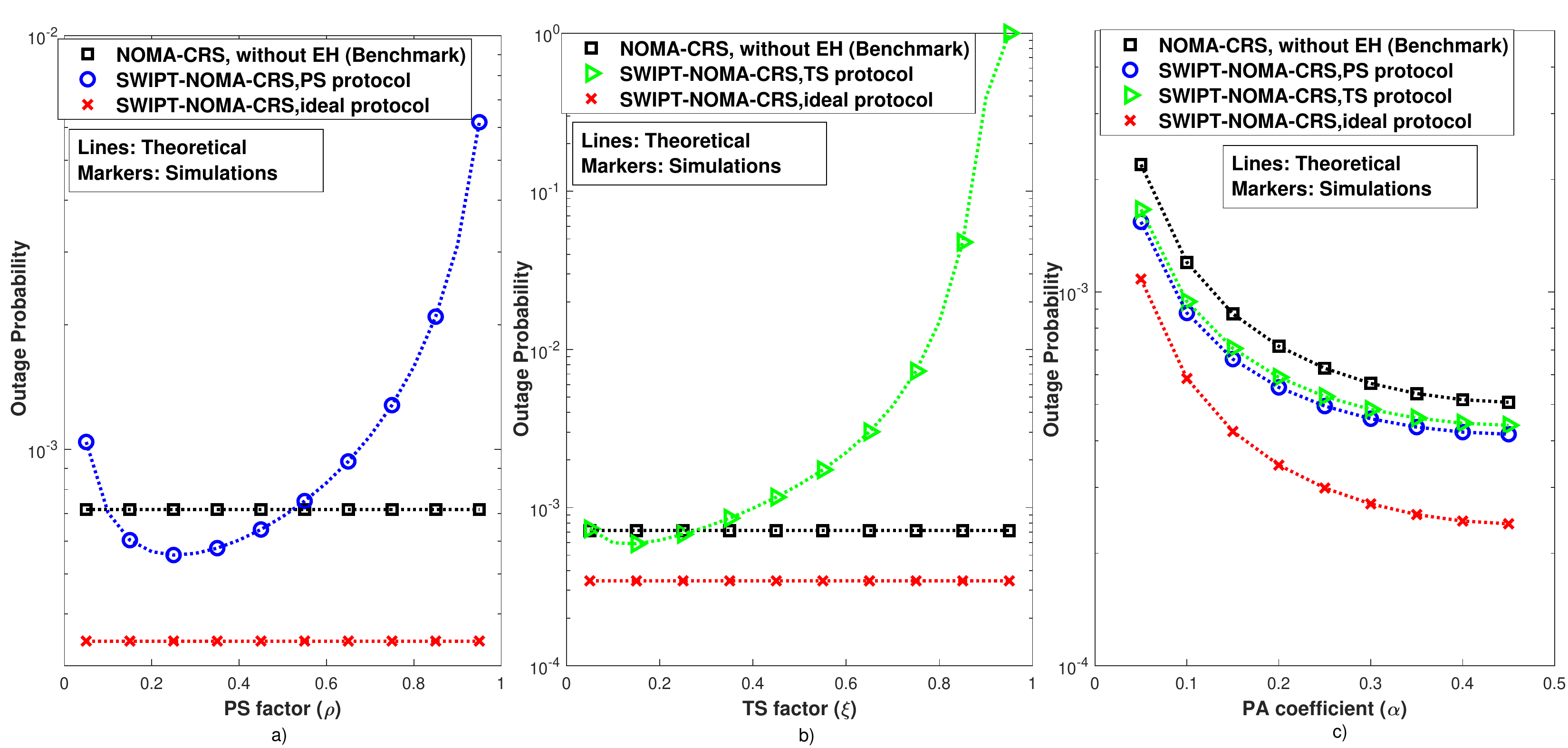}
    \caption{Outage Performance of the SWIPT-NOMA-CRS with respect to EH factor and PA coefficient a) in PS protocol vs. PS factor ($\rho$) b) in TS protocol vs. TS factor ($\xi$) c) in all protocols vs. PA coefficient ($\alpha$) }
    \label{fig7}
 \end{figure}

 In order to reveal the effect of EH protocol and PA parameters, in Figure 7, the OP of the SWIPT-NOMA-CRS is presented with respect to these parameters. In Figure 7, $\delta=0$, $\kappa=0$ and the total SNR is $30\ dB$. In Figure 7(a) and Figure 7(b), the PA coefficient is $\alpha=0.2$. In Figure 7(a), OP of the SWIPT-NOMA-CRS is given in PS protocol with the change of PS factor ($\rho$). To compare, the performances of the ideal protocol in SWIPT-NOMA-CRS and of the NOMA-CRS without EH are also presented. One can easily see that the PS factor ($\rho$) has a dominant effect on the OP and according to chosen $\rho$. The SWIPT-NOMA-CRS with PS protocol can achieve similar performance to ideal protocol or it can have worse performance than NOMA-CRS without EH. To this end, considering the performance gain over NOMA-CRS without EH, the optimum PS factor ($\rho^{*}$) can be given as $0.25$ for the given conditions. Likewise, the OP of the SWIPT-NOMA-CRS in TS protocol is given in Figure 7(b) with respect to TS factor ($\xi$). Based on provided comparisons in Figure 7(b), the SWIPT-NOMA-CRS can outperform NOMA-CRS without EH when only $0.05\leq\xi\leq0.25$ which is a very short range compared to the PS protocol. Then, the OPs of all EH and without EH protocol are presented with respect to PA coefficient ($\alpha$) in Figure 7(c). In Figure 7(c), the PS and TS factors are chosen as $\rho=0.25$ and $\xi=0.15$ according to the obtained values from Figure 7(a) and Figure 7(b). As can be seen from Figure 7(c), with the use of optimal EH factor values, SWIPT-NOMA-CRS always outperforms NOMA-CRS without EH regardless of PA coefficient which is very promising for energy efficiency. With the increase of $\alpha$, the outage performances for all cases becomes better. Nevertheless, this increase is floored after $\alpha\sim0.35$. To this end, for the considered scenario, the optimum PA coefficient can be given as $\alpha^{*}=0.35$.
  \begin{figure}[!t]
   \centering
     \includegraphics[width=16cm]{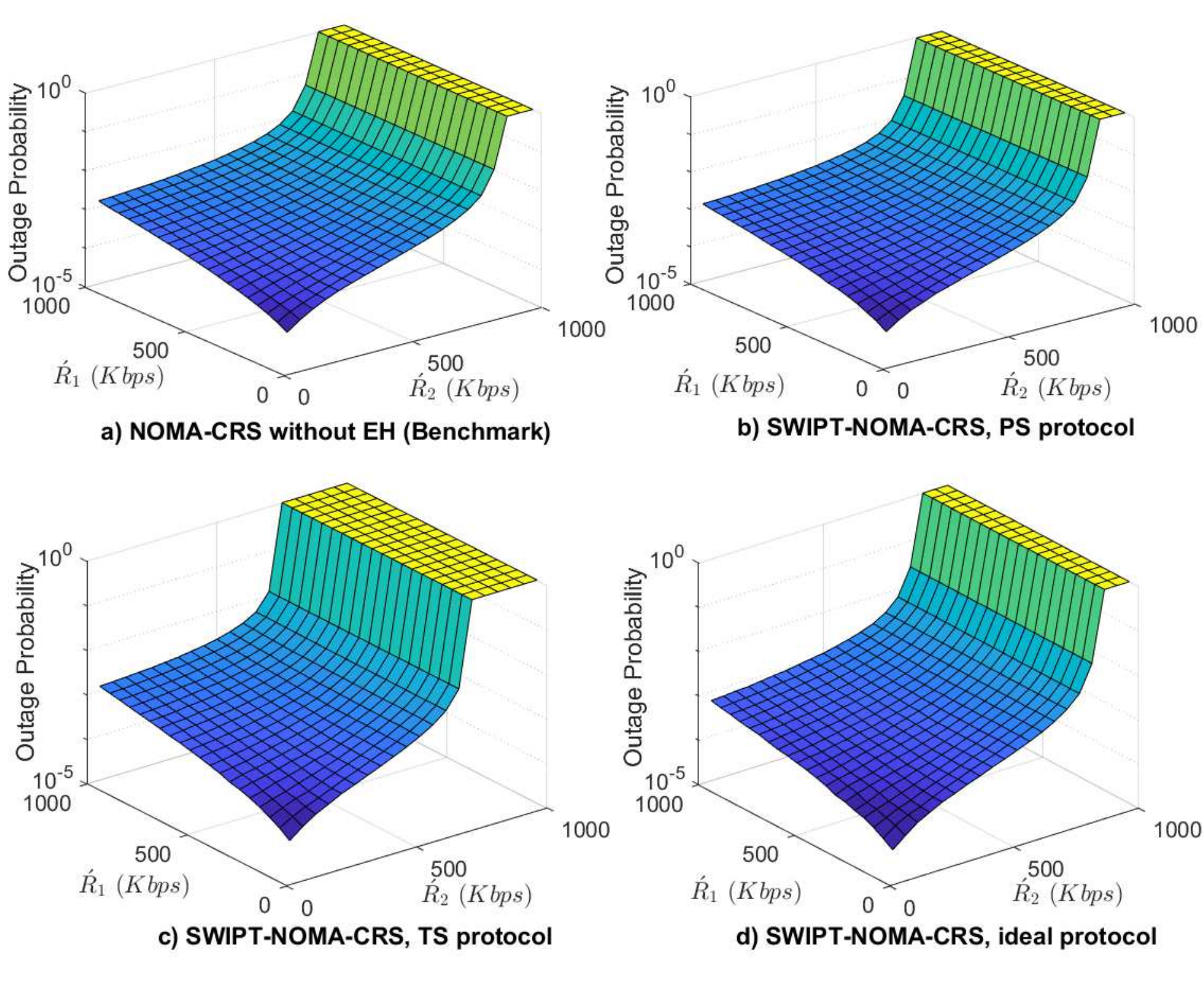}
    \caption{Outage Performance of the SWIPT-NOMA-CRS with respect to target rates ($\acute{R}_1$,$\acute{R}_2$) a) NOMA-CRS without EH (benchmark) b) PS protocol ($\rho=0.25$) c) TS protocol ($\xi=0.15$) d) ideal protocol.}
    \label{fig8}
 \end{figure}

Lastly, to investigate the effects of the target rates (QoS requirements) on the outage performance, in Figure 8, the OPs of the SWIPT-NOMA-CRS are given with respect to $\acute{R}_1$ and $\acute{R}_2$. In Figure 8, the total transmit SNR is $30dB$. The PS factor, TS factor and PA coefficient are set to $\rho=0.25$, $\xi=0.15$ and $\alpha=0.35$ according to discussion on Figure 7. As expected, if one of the target rates is increased, all scenarios have worse outage performance. The best outage performance is achieved when lower rates are required as QoS. On the other hand, if too strict QoS requirements (too high target rates) are demanded, SWIPT-NOMA-CRS with TS protocol may be in always outage. Based on provided comparisons, it is clear that SWIPT-NOMA-CRS outperforms NOMA-CRS without EH for any QoS requirement.

\section{Conclusion}
In this paper, the SWIPT-NOMA-CRS is proposed where the relay node harvests its energy from the RF signal between source and relay. In the SWIPT-NOMA-CRS, three different EH protocol (e.g., PS, TS and ideal protocols) are implemented. The closed-form OP expressions are derived for all EH protocols under imperfect SIC and CSI. The derivations are validated via simulations. The proposed SWIPT-NOMA-CRS outperforms conventional NOMA-CRS without EH significantly and it can reduce the energy consumption up to $\sim5\ dB$ for the same OP target which is very promising for the energy-constraint networks (e.g., IoT). Based on simulations, as expected, the ideal protocol has the best performance. On the other hand, the PS protocol is superior to the TS protocol. Moreover, the effects of the EH and PA parameters on the OP are discussed and it is revealed that the PS protocol is more flexible than the TS protocol. The PS protocols outperforms the conventional NOMA-CRS without EH within a very large PS factor range whereas it is a very small TS factor range in the TS protocol. To this end, the optimum parameters are represented for the minimum OP in given scenarios. Finally, as future works, the energy efficiency of other NOMA schemes can be increased thanks to the SWIPT integration and the analysis of imperfect SIC and CSI can be extended for these systems.


\end{document}

%% file: elksty.tex
\def\E{\ifmmode{\mathbb E}\else{$\mathbb E$}\fi} 
\def\N{\ifmmode{\mathbb N}\else{$\mathbb N$}\fi} 
\def\R{\ifmmode{\mathbb R}\else{$\mathbb R$}\fi} 
\def\Q{\ifmmode{\mathbb Q}\else{$\mathbb Q$}\fi} 
\def\C{\ifmmode{\mathbb C}\else{$\mathbb C$}\fi} 
\def\H{\ifmmode{\mathbb H}\else{$\mathbb H$}\fi} 
\def\Z{\ifmmode{\mathbb Z}\else{$\mathbb Z$}\fi} 
\def\P{\ifmmode{\mathbb P}\else{$\mathbb P$}\fi} 
\def\T{\ifmmode{\mathbb T}\else{$\mathbb T$}\fi} 
\def\SS{\ifmmode{\mathbb S}\else{$\mathbb S$}\fi} 
\def\DD{\ifmmode{\mathbb D}\else{$\mathbb D$}\fi} 

\newcommand{\bse}{\begin{subequations}}
\newcommand{\ese}{\end{subequations}}
\newcommand{\ben}{\begin{enumerate}}
\newcommand{\een}{\end{enumerate}}
\newcommand{\bens}{\begin{enumerate*}}
\newcommand{\eens}{\end{enumerate*}}
\newcommand{\be}{\begin{equation}}
\newcommand{\ee}{\end{equation}}
\newcommand{\bea}{\begin{eqnarray}}
\newcommand{\eea}{\end{eqnarray}}
\newcommand{\baa}{\begin{eqnarray*}}
\newcommand{\eaa}{\end{eqnarray*}}
\newcommand{\bc}{\begin{center}}
\newcommand{\ec}{\end{center}}

\theoremstyle{corollary}

\theoremstyle{lemma}

\theoremstyle{proposition}

\theoremstyle{axiom}

\theoremstyle{conjecture}

\theoremstyle{example}

\theoremstyle{definition}

\theoremstyle{remark}
